\documentclass[aps,twocolumn,groupedaddress,showpacs]{revtex4}
\usepackage{graphicx}
\begin{document}

% Use the \preprint command to place your local institutional report
% number in the upper righthand corner of the title page in preprint mode.
% Multiple \preprint commands are allowed.
% Use the 'preprintnumbers' class option to override journal defaults
% to display numbers if necessary
%\preprint{}

%Title of paper
\title{Binomial states and the phase distribution measurement of weak optical fields}

% repeat the \author .. \affiliation  etc. as needed
% \email, \thanks, \homepage, \altaffiliation all apply to the current
% author. Explanatory text should go in the []'s, actual e-mail
% address or url should go in the {}'s for \email and \homepage.
% Please use the appropriate macro foreach each type of information

% \affiliation command applies to all authors since the last
% \affiliation command. The \affiliation command should follow the
% other information
% \affiliation can be followed by \email, \homepage, \thanks as well.
\author{K. L. Pregnell and D. T. Pegg}
%\email[]{Your e-mail address}
%\homepage[]{Your web page}
%\thanks{}
%\altaffiliation{}
\affiliation{School of Science, Griffith University, Nathan, Brisbane 4111,
Australia}

%Collaboration name if desired (requires use of superscriptaddress
%option in \documentclass). \noaffiliation is required (may also be
%used with the \author command).
%\collaboration can be followed by \email, \homepage, \thanks as well.
%\collaboration{}
%\noaffiliation

\date{\today}

\begin{abstract}
We show that the eight-port interferometer used by Noh, Foug\`{e}res, and Mandel [Phys. Rev. Lett. {\bf 71}, 2579 (1993)] to measure their operational phase distribution of light can also be used to measure the canonical phase distribution of weak optical fields, where canonical phase is defined as the complement of photon number. A binomial reference state is required for this purpose, which we show can be obtained to an excellent degree of approximation from a suitably squeezed state. The proposed method requires only photodetectors that can distinguish among zero, one and more than one photons and is not particularly sensitive to photodetector imperfections.
\end{abstract}

% insert suggested PACS numbers in braces on next line
\pacs{42.50.Dv,  42.50.-p}
% insert suggested keywords - APS authors don't need to do this
%\keywords{}

%\maketitle must follow title, authors, abstract, \pacs, and \keywords
\maketitle

% body of paper here - Use proper section commands
% References should be done using the \cite, \ref, and \label commands

\section{Introduction}

The quantum mechanical nature of the phase of light has been studied since
the beginnings of quantum electrodynamic theory \cite{Dirac}, with renewed
interest recently. The study of quantum phase is distinguished from the
study of many other quantum observables by the difficulties inherent not
only in finding a theoretical description but also in finding methods for
measuring the phase observable so described \cite{Bibliog}. A method of
circumventing this latter difficulty is to define phase as the quantity
measured by some particular experiment. This is known as an operational
approach \cite{B and D}. A sensible operational phase measurement must, of
course, be in accord with a classical phase measurement in the appropriate
limit, for example when the field being measured is in a strong coherent
state. This requirement, however, is not sufficient to define a unique
operational phase observable and various different operational definitions
have been proposed. The best known of these is that of Noh {\it et al.} \cite
{Mandel} who also proposed a means of measuring an operational phase
distribution \cite{Mandel2}, which was further developed in Refs
\cite{Mandel3} and \cite{Mandel4}. There are also various theoretical
approaches
to describing the phase observable. What distinguishes these theoretical
approaches from one another is their predicted phase distributions for
particular states. Some of these approaches are motivated by the aim of
expressing phase as the complement of photon number, in the spirit of
Dirac's original work \cite{Dirac}. Examples of such approaches include the
probability operator measure approach \cite{Hel}, a formalism in which the
Hilbert space is doubled \cite{Newton}, a limiting approach based on a
finite Hilbert space \cite{PB} and a more general axiomatic approach \cite
{Ulf}. Although these particular approaches are quite distinct they all lead
to the same phase probability distribution for a field in state $\left| \psi
\right\rangle $ as a function of phase angle $\theta $ \cite{Ulf}:
\begin{equation}
P(\theta )=\frac{1}{2\pi }\left| \sum_{n=0}^\infty \left\langle \psi
|n\right\rangle \exp (in\theta )\right|^2
\label{1.1}
\end{equation}
where $\left| n\right\rangle $ is a photon number state. Leonhardt {\it et
al. }\cite{Ulf} have called this the ``canonical'' phase distribution to
indicate a quantity that is the canonical conjugate, or complement, of
photon number. Irrespective of how it is derived, the canonical phase
distribution has properties that one might expect from the complement of
photon number: it is shifted uniformly when a phase shifter is applied to
the field, it is not changed by a photon number shift \cite{Ulf} and it
corresponds to a wavefunction in the phase representation of which the
photon number amplitude is the finite Fourier transform \cite
{Bibliog,Conj}. The last property is a natural parallel to
momentum-position conjugacy.

Although the canonical phase distribution has attractive theoretical
properties, its direct measurement presents difficulties \cite{Footnote}.
Good approximate methods exist, based on homodyne techniques, for measuring
the canonical phase distribution of states with narrow phase distributions,
for example coherent states with mean photon numbers of at least five \cite
{VandP}. Weak fields in the quantum regime, however, must have broad phase
distributions because of number-phase complementarity. In principle the
distributions for such fields can be
measured by the projection synthesis method proposed in Ref. \cite{Proj Syn}
but this requires the generation of a reciprocal binomial state as a
reference state. The generation of this exotic state has still not been
achieved. On the other hand the operational phase measurements of Noh{\it \
et al.} are quite practical for the weak fields of interest and have been
shown to measure what they are designed to measure very well \cite{Mandel2}.
Unfortunately they were not designed to measure, nor do they measure, the
canonical phase distribution. Indeed their results show that their
operational phase distribution is significantly different from the canonical
phase distribution. Thus the projection synthesis method measures the
canonical phase distribution in principle while the operational method does
not; but the operational method is practical while the projection synthesis
method is not.

In this paper we show how the apparatus used for measuring the operational
phase distribution can be used to measure the canonical phase
distribution for weak fields in conjunction with a reference field in a
{\it binomial} state. We show how a binomial state, in contrast to a
reciprocal binomial state, can be very well approximated by a squeezed
state, making this method much more practical than the original
projection synthesis method.

\section{projection synthesis}

\subsection{Beam splitter}

In the projection synthesis method of \cite{Proj Syn} the aim was to use
photodetection in conjunction with a beam splitter and a special reference
state to synthesize the projection of an unknown state onto a truncated
phase state
\begin{equation}
\left| \theta \right\rangle =\frac{1}{(M+1)^{1/2}}\sum_{n=0}^M\exp (in\theta
)\left| n\right\rangle
\label{2.1}
\end{equation}
where $M$ must be sufficiently large for the density matrix of the
reproducible weak field that is to be measured to be well approximated in
the number state basis by a matrix with only the first $(M+1)\times
(M+1)$ elements
non-zero. This projection event is associated with the detection of $M$
photons in one output mode of the beam splitter and no photons
in the other. We label this detection event $(M,0)$. The probability of the
event $(M,0)$ is obtained from the occurrence frequency in successive
repeated measurements of the field. The procedure is to measure this
probability as $\theta $ is changed in small steps over the $2\pi $ range.
The changes in $\theta $ can be achieved simply by altering the phase either
of the reference field or of the field to be measured. A histogram is then
plotted which, when suitably normalized, produces the canonical phase
distribution.

The mechanism underlying projection synthesis can be described as follows.
Suppose the measured and reference fields are in the beam splitter
input modes $0$ and $1$
respectively and the event $(M,0)$ is that for which $M$ photons are
detected in output mode $0$ and zero photons are detected in output mode $1$.
The combined output state corresponding to this measurement result is the
$M$-photon state $\left| M\right\rangle _0\left| 0\right\rangle _1$.
Following the unitary evolution of this state backwards through the beam
splitter transforms this to an $M$-photon entangled input state of the form
\begin{equation}
\left| f\right\rangle =\sum_{n=0}^Mf_m\left| n\right\rangle _0\left|
M-n\right\rangle _1
\label{2.2}
\end{equation}
which displays photon number conservation. If the measured field is in a
pure state $\left| \psi \right\rangle _0$ and the reference state is given
by
\begin{equation}
\left| r\right\rangle _1=\sum_{n=0}^\infty r_n\left| n\right\rangle
_1
\label{2.3}
\end{equation}
then the amplitude for the detection event $(M,0)$ is $_{0}\langle \psi
|\, _1\langle r|f\rangle $, that is, the projection of
$|\psi\rangle _0$ onto $_{1}\langle r|f\rangle$. By
choosing appropriate coefficients $r_n$ for $n=0$ to $M$, we can make this
amplitude proportional to the projection of $\left| \psi \right\rangle _0$
onto the truncated phase state $\left| \theta \right\rangle $ given by
(\ref{2.1}). We note that the values of $r_n$ for $n>M$ are irrelevant, merely
affecting the normalization factor for the complete probability
distribution. For a 50:50 symmetric beam splitter we find that the
values of $\left|r_n\right| $ need to be proportional to the reciprocal of
the square root of
the binomial coefficient ${M\choose n}$ for $n=0$ to $M$.

As mentioned earlier, the difficulty with projection synthesis is generating
the reciprocal binomial state required for the reference field. As
reciprocal binomial
states have a finite number of photon number state coefficients, they can be
prepared as travelling fields in principle by the generic methods given in
Refs. \cite{Generation} and \cite{Dakna} by means of beam splitters.
Unfortunately, however, in such techniques the state generated is
conditioned on measuring particular outputs from the beam splitters and so
the method is quite inefficient and difficult and has never been implemented.

\subsection{Multiport device}

Projection synthesis relies on transforming projections onto photon number
states into at least one projection onto a
truncated phase state. As phase is not an absolute quantity, it is
necessary to have a transformation device with at
least two inputs for phase measurement: one for the field to be measured
and one for the reference field. The
projection synthesis method of \cite{Proj Syn} uses the minimal necessary
device, a beam splitter with two inputs and
two outputs. More flexibility can be obtained by using a more general
multiport device with $N+1$ inputs and $N+1$
outputs with a photodetector in each output. This would require, in
addition to the field in state $|\psi\rangle_0$ to
be measured being in input mode $0$, say, $N$ reference fields being in
modes 1,2 $\ldots N$. Rather than exacerbate
the problem of preparing special reference states, it is preferable simply
to have one reference field in input mode 1
with vacuum fields in the remaining inputs. These input states are
transformed by a unitary operator $\widehat{R}$ into
the output states. If a total of $M$ photons are detected in the output
states then, from photon number conservation,
the corresponding $M$-photon output state evolved backwards through the
multiport device will be transformed to an
entangled $M$-photon input state. Projecting the combined vacuum input
state $\left| 0\right\rangle _2\left|
0\right\rangle _3\ldots \left| 0\right\rangle _N$ onto this entangled state
will result in a two-mode $M$-photon
entangled state in input modes $0$ and $1$ of the form $\left|
f\right\rangle $ in (\ref{2.2}). The coefficients $f_m$
will depend on the unitary transformation $\widehat{R}$ and on the manner
in which the $M$ photons are detected, that
is, how they are distributed over the output modes. As these coefficients
determine the values of $r_n$ required to
synthesize the projection of the state to be measured onto a truncated
phase state, use of a multiport device should
increase the flexibility in choosing a convenient reference state
$|r\rangle _1$.

A natural multiport extension of the 50:50 symmetric beam splitter is one
where $\widehat{R}$ is such that a photon entering any input appears with
equal probability at any output \cite{Mattle,Torma} and a photon in any
output is equally likely to have come from any input. In this case it is not
difficult to show that, when the $M$ photons are detected in output mode $0$
and none are detected in the other modes, the coefficients $f_m$ are similar
to those for the beam splitter case for the event $(M,0)$ and we again
require the reference field to be in a reciprocal binomial state. Thus it is
worth examining other possible distributions of the detected photons over
the output modes.

A particular case of the above device is one that performs a discrete
Fourier transform \cite{Torma}, that is where the
set of mode photon creation operators and the set of transformed operators
form a discrete Fourier transform pair:
\begin{equation}
\widehat{R}^{\dagger }\widehat{a}_j^{\dagger }\widehat{R}=
\sum_{i=0}^NU_{ij}\widehat{a}_i^{\dagger }
\label{2.4}
\end{equation}
\begin{equation}
\widehat{a}_i^{\dagger }=\sum_{j=0}^NU_{ij}^{*}
\widehat{R}^{\dagger}\widehat{a}_j^{\dagger }\widehat{R}
\label{2.5}
\end{equation}
where
\begin{equation}
U_{ij}=\frac{\omega ^{ij}}{\sqrt{N+1}}\label{2.6}
\end{equation}
with $\omega =\exp[-i2\pi/(N+1)]$, that is, a $(N+1)$th root of unity.
Recently \cite{PregPRL}, we found that, if the
number $M$ of photons detected is equal to $N$ and these photons are
distributed such that one is detected in each
output of such a multiport device except one, then projection onto a
truncated phase state is synthesized providing the
reference field is in a {\it binomial} state. Specifically, if the
detection event is such that it is the $m$th output
detector that records a zero count while all the other detectors record one
count each, then the state $|\psi\rangle_0$
to be measured is projected onto a state proportional to
\begin{equation}
\left| \theta _m\right\rangle_0 =\sum_{n=0}^NU_{mn}^{*}\left|
n\right\rangle_0 \label{2.7a}
\end{equation}
that is
\begin{equation}
\left| \theta _m\right\rangle_0 =\frac 1{(N+1)^{1/2}}\sum_{n=0}^N\exp
(in\theta _m)\left| n\right\rangle_0 \label{2.7b}
\end{equation}
where
\begin{equation}
\theta _m=m2\pi /(N+1)\label{2.8}
\end{equation}
which is just the required truncated phase state. The amplitude for this
detection event is proportional to
${}_0\langle\theta_m|\psi\rangle_0$ \cite{PregPRL}.

In addition to requiring photodetectors that only need to distinguish among
zero photons, one photon and more than one
photon, this multiport device has the advantage that the required binomial
reference state is much closer to commonly
available states such as coherent states than is a reciprocal binomial
state. Binomial states have been studied for
some time \cite{Stoler}. They have interesting properties such as
interpolating between coherent states and number
states and, with the photon number state coefficients positive, they are
partial phase states with a mean phase of zero
\cite{Vid}. It is not difficult to show that they have a smaller phase
variance than truncated phase states with the
same number of number state coefficients. As binomial states have a finite
number of photon number state coefficients,
they can be prepared as travelling fields by the generic methods given in
Refs. \cite{Generation} and \cite {Dakna} by
means of beam splitters. Unfortunately, however, these inefficient
techniques offer no real advance when used for
measuring the phase distribution over the use of a reciprocal binomial
state in \cite{Proj Syn}, which can be prepared
by the same generic means. For a practical experiment we require the
binomial state to be approximated by a state which
is reasonably straightforward to prepare on demand.

As mentioned earlier, the values $r_n$ of the number state coefficients of
the reference state for $n>M$ are irrelevant
so here, where $M=N$, only the coefficients $r_n$ with $0\leq n\leq N$ are
important, and thus only these need to be
proportional to square roots of binomial coefficients. Further, as we are
interested in fields with broad phase
distributions, and states with broad number state distributions tend to
have narrow phase distributions, then normally
only a small group of number state coefficients $_0\langle n|\psi\rangle
_0$ of the state of the measured field will
differ significantly from zero. From (\ref{2.2}) with $M=N$ and
(\ref{2.3}), if the coefficient $_0\langle n|\psi
\rangle _0$ is significant then the value of $r_{N-n}$ is important. Thus,
for example, if $_0\langle n|\psi\rangle _0$
are significant only for a small number of values of $n$ equal to or
slightly less than a value $n^{\prime }$, say,
then we would choose $N=n^{\prime }$ and require the small number of
coefficients $r_n$ with $n$ equal to, or near,
zero to be proportional to square roots of appropriate binomial
coefficients. We show in the Appendix how such a state
can be approximated by a squeezed state with squeezing parameter $\tanh
^{-1}0.5$. On the other hand if the significant
values of $_0\langle n|\psi\rangle_0$ occur for $n$ equal to or near zero,
as will be the case for very weak fields,
then we require coefficients $r_n$ with $n$ equal to or slightly less $N$
to be proportional to square roots of the
appropriate binomial coefficients. In the Appendix we find a squeezed
state, also with squeezing parameter of $\tanh
^{-1}0.5 $, that is a very good approximation for the binomial reference
state needed for measuring the phase
distribution of very weak fields with the eight-port interferometer
examined below.

\section{Eight-port interferometer}

\subsection{Binomial reference state}

\begin{figure}
\includegraphics[width=0.40\textwidth]{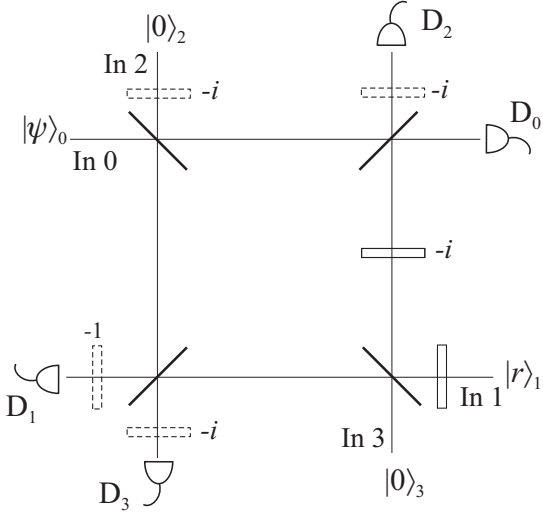}
\caption{Eight-port interferometer for measuring the canonical phase
distribution of weak fields. The field in state
$|\psi\rangle_0$ to be measured is in the input port labelled In 0, the
reference field in state $|r\rangle_1$ is in
input port In 1 and vacuum state fields are in In 2 and In 3. A
photodetector is in each output port. The dotted phase
shifters are for mathematical convenience only, and do not affect the
results.} \label{fig1}
\end{figure}

The eight-port interferometer \cite{Walker} used by Noh {\it et al. }\cite
{Mandel,Mandel2}{\it \ }and Torgerson and
Mandel \cite{Mandel2,Mandel3} is illustrated in Fig. \ref{fig1}. There are
four 50:50 symmetric beam splitters at the
corners of a square. The phase shifter labelled $-i$ between the two beam
splitters on the right shifts the phase by
$\pi /2$. The field state $|\psi\rangle_0$ to be measured is in input port
0. The phase shifter in input port $1$
allows the phase of the reference field state $|r\rangle_1$ to be changed.
The dotted phase shifters in input port $2$
and before detectors D$_{1}$, D$_2$ and D$_3$, which are not present in the
original interferometer, are merely
inserted here for mathematical convenience. As the field in input port 2 is
the vacuum it will not be affected by a
phase shift and, as the detectors detect photons, their operation will not
be affected by phase shifters in front of
them.

A single 50:50 symmetric beam splitter transforms the input creation
operators $\widehat{b}^{\dagger }$ and $\widehat{c}^{\dagger }$ in accord
with \cite{Book}
\begin{equation}
\widehat{R}_1\widehat{b}^{\dagger }\widehat{R}_1^{\dagger }=2^{-1/2}(%
\widehat{b}^{\dagger }+i\widehat{c}^{\dagger })\label{2.9}
\end{equation}
\begin{equation}
\widehat{R}_1\widehat{c}^{\dagger }\widehat{R}_1^{\dagger }=2^{-1/2}(%
\widehat{ib}^{\dagger }+\widehat{c}^{\dagger })\label{2.10}
\end{equation}
where $\widehat{R}_1$ is the unitary operator for the action of the single
beam splitter. By using this relation successively, it is not difficult to
show that the input creation operators for the eight-port interferometer,
including the dotted phase shifters, are transformed as
\begin{equation}
\widehat{R}\widehat{a}_i^{\dagger }\widehat{R}^{\dagger }=\exp (i\gamma
)\sum_{i=0}^3U_{ij}^{*}\widehat{a}_j^{\dagger }\label{2.11}
\end{equation}
where
\begin{equation}
U_{ij}=\frac{\omega ^{ij}}2\label{2.12}
\end{equation}
with $\omega$ again being $\exp[-i2\pi/(N+1)]$, provided we set the phase
shifter in input $1$ to shift the phase by
$\pi /2$, that is to attach a value $-i$ to it. Expressions (\ref{2.11})
and (\ref{2.12}) are in agreement with
(\ref{2.5}) and (\ref{2.6}) for $N=3$ apart from the phase factor $\exp
(i\gamma )$, which depends on the difference
between the distance between beam splitters and an integer number of
wavelengths. This phase factor does not affect the
photocount probabilities and can be ignored.

Thus we see that the eight-port interferometer, without modification, can
be used to synthesize the projection of the state to be measured onto one of
four phase states. Specifically the probability of measurement event (0, 1,
1, 1), that is the detection of zero photocounts in detector D$_0$ and one
in each of D$_1$, D$_2$ and D$_3$, is proportional to the square of the
modulus of the projection of the measured state onto the truncated phase
state
\begin{equation}
\left| \theta _0\right\rangle =2^{-1}(\left| 0\right\rangle +\left|
1\right\rangle +\left| 2\right\rangle +\left| 3\right\rangle )\label{2.13}
\end{equation}
while the probability of the event (1,0,1,1) is proportional the square of
the modulus of the projection of the measured state onto the truncated phase
state

\begin{equation}
\left| \theta _1\right\rangle =2^{-1}(\left| 0\right\rangle +i\left|
1\right\rangle -\left| 2\right\rangle -i\left| 3\right\rangle )\label{2.14}
\end{equation}
and so on, in accord with (\ref{2.7b}) with $N=3$. Repeating the experiment
a number of times with a reproducible state
will allow a probability $P_{M}(\theta_{m})$ with $m=0,1,2,3$ to be
measured for each of the four events (0,1,1,1),
(1,0,1,1), (1,1,0,1) and (1,1,1,0) respectively.

To use these four measured probabilities to construct the phase distribution
we first normalize them to
\begin{equation}
y(\theta _m)=\frac{2P_{M}(\theta_{m})}{\pi\sum_m P_{M}(\theta _m)}.\label{2.15}
\end{equation}
We note that, as shown in general in Ref. \cite{PregPRL},
$P_{M}(\theta_{m})\propto
|\langle\psi|\theta_{m}\rangle|^{2}$, where here $|\theta _m\rangle$ is
given by (\ref{2.7b}) with $N=3$ and we have
omitted the subscripts zero for convenience. This allows us to replace
$P_{M}(\theta_{m})$ in (\ref{2.15}) by
$|\langle\psi|\theta_{m}\rangle|^{2}$. Substituting for $\theta_m $ from
Eq. (\ref{2.8}) with $N=3$ yields eventually
\begin{equation}
\sum_{m=0}^3|\langle\psi|\theta_{m}\rangle|^{2}=| \langle \psi |0\rangle
| ^2+| \langle \psi |1\rangle | ^2+|
\langle \psi |2\rangle | ^2+|\langle \psi
|3\rangle | ^2.\label{2.16}
\end{equation}
If the measured field is sufficiently weak for the number state components
$\left\langle \psi |n\right\rangle $ to be negligible for $n\geq 4$ then the
right-hand side of Eq. (\ref{2.16}) is just unity and so, from (\ref{2.15}),
\begin{eqnarray}
y(\theta_{m})&=&\frac{2}{\pi}|\langle\psi|\theta_{m}\rangle|^{2}\nonumber
\\&=&P(\theta _m),\label{2.17}
\end{eqnarray}
with the last line being obtained from Eq. (\ref{1.1}) in the same
weak-field approximation.

Thus, if we normalize the four measured probabilities by dividing each by
the sum of the four and then multiplying by $2/\pi $, we obtain four points
on the canonical phase probability distribution given by Eq. (\ref{1.1}) for a
weak field. We note that, with this normalization procedure, if the points
$y(\theta _m)$ are used to draw a histogram, the area of the histogram would
unity, because the width of each rectangle is $\pi /2$.

Shifting the phase of the phase shifter in the input $1$ to change the phase
of the reference field by $\Delta \theta $ and repeating the procedure gives
four more points of the distribution shifted from the original points by $%
\Delta \theta $. A sixteen point curve, for example, can be constructed by
shifting the phase by $\pi /8$ three times and repeating the experiment
after each shift.

\subsection{Squeezed reference state}

\begin{figure}
\includegraphics[width=0.40\textwidth]{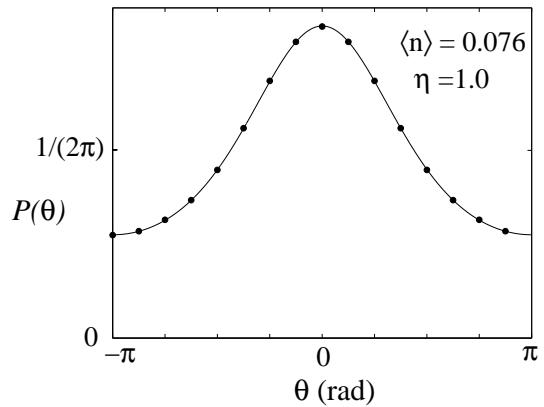}
\caption{Canonical phase probability distribution $P(\theta)$ for a
coherent state field with a mean photon number of
0.076. The full line is the theoretical result and the dots are simulated
measured results with ideal detectors for
four different phase settings of the squeezed reference field.} \label{fig2}
\end{figure}

\begin{figure}
\includegraphics[width=0.40\textwidth]{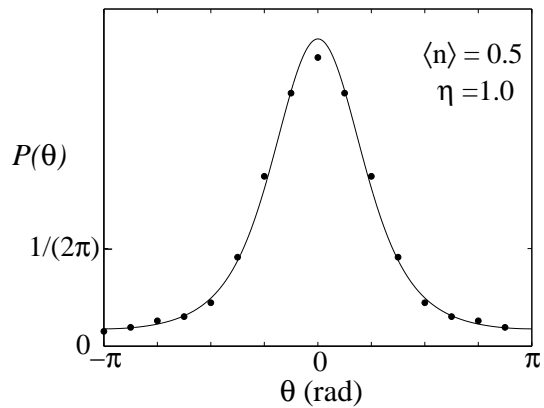}
\caption{Simulated measurements (dots) and the theoretical
canonical phase distribution (full line) for a coherent state field
with a mean photon number of 0.5 with ideal detectors.}
\label{fig3}
\end{figure}

The above analysis and suggested procedure assumes that the reference field
is in a perfect binomial state. If, instead, we use the squeezed state
approximation to the binomial state as derived in the Appendix, then
the vacuum state coefficient differs from the ideal value and the measured
state is no longer projected onto the truncated phase state $\left| \theta
_m\right\rangle $ but is instead projected onto a state proportional to
\begin{equation}
| 0\rangle +\exp (i\theta _m)| 1\rangle +\exp
(2i\theta _m)| 2\rangle +1.0146\exp (3i\theta _m)|
3\rangle. \label{2.20}
\end{equation}
We would expect that this would lead to some small errors when the procedure
suggested above is applied. In practice, if we are measuring a state, such
as a coherent or squeezed state, that does not have a truncated photon
number distribution, the error caused by the modulus of the
three-photon
coefficient in expression (\ref{2.20}) differing from unity may in
general be smaller that the error caused by the truncation of expression
(\ref{2.20}) after the three-photon
component. In Fig. \ref{fig2} we show the points obtained from a simulated
experiment for a coherent state with a mean photon number of 0.076, which
is is comparable to the field strength of interest in Ref. \cite{Mandel3},
using a squeezed reference state. The close agreement with the
canonical distribution is apparent. For weaker fields, for example the other
field of interest in Ref. \cite{Mandel3} with a mean photon number of 0.047,
the agreement is even closer. Agreement is still good for stronger coherent
state fields with mean photon numbers of 0.139 and 0.23, as used in Ref.
\cite{Mandel4}, with divergence from the canonical distribution becoming
apparent for mean photon numbers of around 0.4. Figure \ref{fig3} shows
simulated
results for a coherent state with a mean photon number of 0.5. The error
here is almost entirely due to the truncation of the phase state rather than to
the non-unit coefficient of the fourth term in (\ref{2.20}). A mean photon
number of 0.5
represents the approximate limit to the field strength for a coherent state
for which this measurement technique is suitable.

\section{Some practical considerations}

To obtain an idea of the number of experiments needed to measure the
required probabilities, we note that, for weak coherent fields of the
strengths discussed here, the coefficient of the vacuum component dominates,
so the probability of detecting a total of three photons will be
approximately the probability that the binomial reference state contains
three photons, which is 12.5\%. These photons can be detected as a variety
of events such as (0,1,1,1), (0,0,2,1) or (0,0,0,3). The fraction of these
events that are required events, that is three separate counts of one photon
and one count of zero, is 3/32. Thus approximately 4.7\% of experiments will
produce one of the four desired events. After running the experiment
sufficient times to obtain the four probabilities with the desired accuracy,
the phase of the reference state is then changed and the procedure is
repeated to obtain four more probabilities. The question therefore arises:
how many probabilities, that is points on the probability distribution, do
we need to determine the distribution with reasonable accuracy? The fact that
we are synthesizing a projection onto a truncated phase state with four
non-zero photon number-state coefficients means that the weak fields being
measured must, by necessity, have negligible coefficients for components
with photon numbers greater than three. From Eq. (\ref{1.1}) this means
that the
most rapidly oscillating Fourier component of the distribution $P(\theta )$
behaves as $\exp (i3\theta )$. To detect this oscillation we would need
a minimum of about 12 equally spaced values of $P(\theta )$. Thus running
the experiment with four different phase settings, giving a total of
16 points on the distribution curve as shown in Fig. 2, should normally
be sufficient.

In a practical experimental situation errors can arise from collection
inefficiencies, non-unit quantum efficiencies for one, two and multiple
photon detection, dead times and accidental counts  arising from dark counts
and background light. The fact, however, that
Noh {\it et al.} \cite{Mandel} have performed successful
experiments involving the measurement of joint detection probabilities
with an eight-port interferometer, by means of photon counting,
for states with field strengths similar to those of interest in this paper
is an encouraging indication that there should be no insurmountable
difficulties for the method proposed here arising from such errors. It is
worth considering some specific aspects of the sources of error.
In the experiments of Noh {\it et al.} \cite{Mandel} photon count rates
were of the order of $10^4$ per second with a counting interval of
about 5 $\mu$s, to give the required small mean photon number, and
dead-time effects were negligible.  In the present proposed experiment
dead times are even less important because it is only necessary to
discriminate among zero, one and many counts rather than among
general numbers of counts \cite{Paul}.  Dark counts can be reduced  to
about 200 per second \cite{Mandel} or even to 20 per
second \cite{Trif} by cooling the detectors and background light can be
reduced
by appropriate shielding.  In the event that the residual dark and
background counts are not negligible, the measured joint probabilities
of the four photocount events can be corrected by a deconvolution
procedure using the data obtained by blocking the input
signals \cite{Mandel}.

Concerning detector efficiencies, even if collection efficiencies
are made to approach unity by, for example, suitable geometry and
reflection control,
there will still be some detector inefficiency due to non-unit
quantum efficiency, so some correction for detector
inefficiency may be needed. Conventional
single-photon counting module detectors can have an efficiency of
around 0.7 \cite{Tak}, while visible light photon counters that
distinguish between single-photon and two-photon incidence can have
quantum efficiencies of about 0.9 with some sacrifice of smallness of
dark count rate \cite{Tak}.
We denote the one-photon detection
efficiency, that is the probability of recording one photocount if one
photon is present, by $\eta$. Then, as dead times are not important, the
general multiple detection efficiency is such that the probability of
recording $n$ photocounts if $N$ photons are present is
${N\choose n}\eta^{n}(1-\eta)^{N-n} $ where the
first factor is the binomial coefficient \cite{Lee}. If $\eta$ is the same
for all four detectors the probability for the joint four-count
detection event $(m,n,p,q)$ is given by
\begin{eqnarray}
    P_{c}(m,n,p,q)=\sum_{s=m}^{\infty}\sum_{t=n}^{\infty}
    \sum_{u=p}^{\infty}\sum_{v=q}^{\infty} {s\choose m}{t\choose n}
    {u\choose p}{v\choose q}\nonumber
    \\ \times
    \eta^{m+n+p+q}(1-\eta)^{s+t+u+v-m-n-p-q}P_{I}(s,t,u,v)\label{2.21}
\end{eqnarray}
where $P_{I}(s,t,u,v)$ is the probability that an ideal
detector would have detected the joint four-count event $(s,t,u,v)$.
The relation (\ref{2.21}) can be inverted by use of the four-function
Bernoulli transform, which is straightforward to derive in a similar
manner to that of the two-function transform derived in Ref. \cite{PeB}.
This allows us to calculate the ideal probabilities from the
measured probabilities and thus correct for non-unit efficiencies.

\begin{figure}
\includegraphics[width=0.40\textwidth]{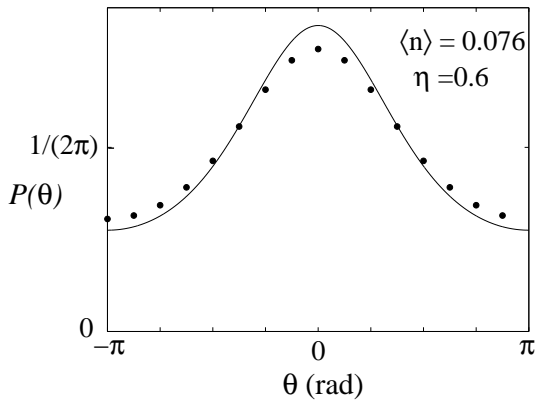}
\caption{Uncorrected simulated measurements (dots) and the theoretical
canonical phase distribution (full line) for a coherent state field
with a mean photon number of 0.076 where the photodetectors have an efficiency
$\eta = 0.6$.}
\label{fig4}
\end{figure}

Although we can correct for non-unit efficiencies an analysis
shows that the effect of not correcting for them is not as serious
as it may first appear.
Essentially this is because the four measured probabilities are always
normalized so their sum is $2/\pi $. The major effect of $\eta $ not being
unity is, as can be seen from (\ref{2.21}), that the probabilities for
the four events (0,1,1,1), (1,0,1,1),
(1,1,0,1) and (1,1,1,0) to be actually recorded are reduced by a factor
$\eta ^3$. As this affects the event probabilities uniformly, however, the
effect vanishes upon normalization. The next order effect is that some ideal
four-count events, such as (1,1,1,1) and (0,2,1,1), are registered, for
example, as (0,1,1,1) because of the inefficiency. The effect of this is
only partially removed by the normalization. Ideal higher-count events also
contribute to the error, but for weak fields the probability of ideal
high-count events is not large. A numerical calculation of the total effect
of non-unit efficiency, including the effect of normalization, shows that
the proposed procedure is not highly sensitive to
detector inefficiency, provided the efficiency is reasonable, for the weak
states of interest. More precisely, for coherent states with a mean photon
number up to 0.5 photons, as discussed above, the error in the final
normalized probabilities is less than 2\% for $\eta \geq 0.9$. For a mean
photon number of 0.076, the error is less that 0.5\% for such efficiencies.
In Fig. \ref{fig4} we show the effect of a poorer efficiency of $\eta =0.6$
for a
mean photon number of 0.076.

To produce the squeezed state required as an approximate binomial state, we
note that squeezed vacuum states can be transformed into various types of
squeezed states, the squeezing axis can be rotated, coherent amplitude can
be added and the squeezing can be controlled independently of the coherent
amplitude. The degree of squeezing needed here is of a magnitude that is a
realistic expectation either now or in the near future \cite{Bachor}.

Our discussion has focused mainly on measuring the phase
probability distribution of pure states of light. It is
straightforward, however, to show that our procedure can also be used
to measure the distribution of a mixed state. If the
mixed state is represented by a density operator that is a weighted
sum of pure state density operators $|\psi\rangle \langle\psi|$ for
various states $|\psi\rangle$, then the phase probability density
is just the corresponding weighted sum of values of $P(\theta)$ given
expression (\ref{1.1}) for the various states $|\psi\rangle$. To
measure this, we can use precisely the same procedure as described in this
paper with the same binomial reference state.

\section{conclusion}

We have shown in this paper how the eight-port interferometer used by Noh
{\it et al.} \cite{Mandel,Mandel2} to measure their operational phase
distribution can used to measure the canonical
phase distribution given by Eq. (\ref {1.1}), where the canonical phase is
defined as the complement
of photon number. The procedure is applicable for weak fields in the quantum
regime, by which we mean explicitly states for which number state components
for photon numbers greater than three are negligible. For coherent states,
this requirement translates to a mean photon number of a half a photon or
less. This is precisely the quantum regime in which large differences
between the operational phase and the canonical phase distributions are most
apparent. For example fields of interest in Refs \cite{Mandel3,Mandel4} are
coherent states with mean photon numbers of 0.23, 0.139, 0.076 and 0.047.
The success of the experiments in the foregoing references indicates that
the procedure proposed in this paper should be viable, given a reliable
source of the required reference state.

The procedure in this paper has advantages over the original projection
synthesis method proposed for measuring the canonical phase distribution.
The most significant of these is that the present procedure requires a
binomial reference state rather than a reciprocal binomial state. We have
shown that the required binomial state can be approximated by a squeezed
state sufficiently well for our purpose. Another advantage is that we
require only photodetectors that can distinguish among zero, one and more
than one photocounts. The measurements are not particularly sensitive to
photodetector inefficiency and, for reasonably good detector efficiencies,
no corrections should be needed. Overall, we feel that the proposal in this
paper brings the measurement of the canonical phase distribution for weak
optical fields closer to reality.

\appendix*
\section{Binomial states}

In this Appendix we show how the required binomial reference state
can be approximated by a suitably squeezed state. The particular
binomial state of interest to us is given by
\begin{equation}
\left| B\right\rangle =\sum_{n=0}^N\,\beta _n\left| n\right\rangle
=2^{-N/2}\sum_{n=0}^N\,{N\choose n}^{1/2}\left| n\right\rangle \label{A1}
\end{equation}
where ${N\choose n}$ is the binomial coefficient. The binomial state
derived in Ref. \cite{PregPRL} with alternating signs for the number
state coefficients can be obtained by phase shifting this state by
$\pi$.

The general form for a squeezed state is \cite{Loudon}
\begin{eqnarray}
\left| \alpha ,\zeta \right\rangle &=&\sum_{n=0}^\infty \,\alpha _n\left|
n\right\rangle \nonumber
\\&=&(\cosh |\zeta |)^{-1/2}\exp \{-\frac 12[|\alpha |^2+
t(\alpha ^{*})^2]\} \nonumber
\\
&&\times \sum_{n=0}^\infty \frac{(t/2)^{n/2}}{(n!)^{1/2}}H_n\left[\frac{\alpha
+t\alpha ^{*}}{(2t)^{1/2}}\right]\left| n\right\rangle
\label{A2}
\end{eqnarray}
where $\zeta =|\zeta |\exp (i\phi )$ with $|\zeta |$ being the squeezing
parameter, $t=\exp(i\phi )\tanh |\zeta |$ and $H_n(x)$ is a Hermite
polynomial of order $n$. $\alpha$ is the complex amplitude of the coherent
state obtained in the limit of zero squeezing.

The first case we study is where we are interested in finding a
squeezed state whose coefficients $\alpha _n$ are proportional to the
coefficients $\beta _n$ of binomial state for the early terms, that is for
$n<<N$. In this case we can approximate the binomial coefficient by
\begin{eqnarray}
{N\choose n}^{1/2} &=&\frac{N^{n/2}}{\sqrt{n!}}\sqrt{(1-\frac{1}{N})
(1-\frac{2}{N})\ldots (1-\frac{n-1}{N})} \nonumber
\\
&\approx &\frac{N^{n/2}}{\sqrt{n!}}\left[ 1-\frac{n(n-1)}{4N}\right]
\label{A3}
\end{eqnarray}
We can approximate the Hermite polynomial for large $x$ by its leading
terms:
\begin{equation}
H_n(x)\approx (2x)^n-n(n-1)(2x)^{n-2}\label{A4}
\end{equation}
We find remarkably that choosing $t=0.5$ and $\alpha =(2/3)N^{1/2}$ allows
us to write
\begin{equation}
{N\choose n}^{1/2}\approx \frac{(t/2)^{n/2}}{(n!)^{1/2}}H_n\left[\frac{\alpha
+t\alpha ^{*}}{(2t)^{1/2}}\right] \label{A5}
\end{equation}
for $n<<N$. Thus the first $n$ number state coefficients of a
squeezed state with these values of $t$ and $\alpha$ will be
proportional to the required binomial coefficients to a good
approximation. With this degree of squeezing, the squeezed quadrature variance
is 1/3 that of the vacuum level, that is, 4.77 dB below the standard
quantum limit.

The opposite case to the above is where we require a small number of
coefficients $\alpha _n$ for $n=N,N-1,N-2\ldots $ to be proportional to
$\beta _n$. It is not as easy to obtain as general a relationship as the
above so we look at each case individually. In this paper we are interested
in the particular case with four values of $\beta _n$, that is, $N=3$. By
using the explicit form of the Hermite polynomials in Eq. (\ref{A2})
and setting $\alpha
_2/\alpha _3=\beta _2/\beta _3$ and $\alpha _1/\alpha _3=\beta _1/\beta _3$
we find that the values $t=0.5$ and $\alpha =(2+2^{1/2})/3$ satisfy the two
simultaneous equations obtained. We note that the required squeezing
parameter $\tanh ^{-1}0.5$ is the same as for the first case above but the
value 1.138 of $\alpha $ varies slightly from 1.155, the value of $%
(2/3)N^{1/2}$ with $N=3$, which is required to make the first few
coefficients of $\left| \alpha ,\zeta \right\rangle $ proportional to
binomial coefficients. We also note that with perfect matching of the last
three coefficients the ratio $\alpha _0/\alpha _3$ becomes 1.0146, a
mismatch of only 1.5\% with the corresponding binomial coefficient.

% If you have acknowledgments, this puts in the proper section head.
\begin{acknowledgments}
D. T. P. thanks the Australian Research Council for funding.
\end{acknowledgments}

% Create the reference section using BibTeX:
%\bibliography{basename of .bib file}

\end{document}